\documentstyle[preprint,aps]{revtex}

\tightenlines

\begin{document}

\draft

\title{
An estimate of a lower bound on the masses of mirror baryons
}

\author{
A.~Garc\'{\i}a
}
\address{
Departamento de F\'{\i}sica.\\
Centro de Investigaci\'on y de Estudios Avanzados del IPN.\\
A.P. 14-740. M\'exico, D.F., 07000. MEXICO.
}
\author{
R.~Huerta
and
G.~S\'anchez-Col\'on
}
\address{
Departamento de F\'{\i}sica Aplicada.\\
Centro de Investigaci\'on y de Estudios Avanzados del IPN. Unidad Merida.\\
A.P. 73, Cordemex. M\'erida, Yucat\'an, 97310. MEXICO.
}

\date{\today}

\maketitle
 
\begin{abstract}
We consider the most favorable conditions to indirectly observe the mixing of
ordinary and mirror hadrons in non-leptonic and weak radiative decays of
hyperons. This allows us to set a lower bound on the masses of mirror baryons.
This bound turns out to be impressively high, of the order of $10^6$GeV.
\end{abstract}


In this paper we shall obtain a lower bound on the masses of mirror hadrons
under what are probably the most favorable conditions to indirectly observe
such hadrons in low energy physics. Lee and Yang in their pioneering
paper~\cite{lee} about parity violation in nature discussed in detail the
existence of mirror matter as a possible framework to help us understand such
violation. Ever since, mirror matter has been discussed at different levels,
including extensions of the minimal standard model~\cite{mohapatra}. Not only
hadrons and leptons (or quarks and leptons) might have mirror partners but the
electroweak gauge bosons too would have mirror partners. The photon might be
accompanied by a massless paraphoton, which however could not couple to
ordinary fermions~\cite{glashow}. Extensions of the standard model doubling the
electroweak gauge group and the fermion content have been discussed in detail
in the literature~\cite{barr}. Such extensions might provide a solution to the
problem of strong violation of $CP$. Of all the possibilities considered the
most attractive one and still in the spirit of Ref.~\cite{lee}, which we shall
refer to as manifest mirror symmetry, is the one in which both ordinary matter
(hadrons and leptons) and mirror matter (mirror hadrons and leptons) share the
same strong and electromagnetic interactions. Other alternatives, even if they
keep the same particle content, are intuitively less attractive.

Manifest mirror symmetry opens the possibility that mirror hadrons become
observable in low energy physics through their mixing with ordinary hadrons.
The reason for this is that, since they share strong and electromagnetic
interactions with ordinary hadrons, even if the mixing angles are very small
they still could lead to observable effects that might compete with weak
interactions process. In previous papers~\cite{cinco,seis} we studied how such
mixing might mimic non-leptonic and weak radiative decays of hyperons (NLDH and
WRDH, respectively). This mechanism, which we referred to as a priori mixing,
might explain the $|\Delta I|=1/2$ rule observed in such decays, in case (for
some as yet unknown reason) the enhancement of the $W$-boson contributions to
NLDH and WRDH could not be produced within the minimal standard model. We
remind the reader that despite the effort invested in this direction no final
answer to this problem has been produced so far, although of course the
possibility of obtaining it remains quite open. If this last were to be the
case then the data on NLDH and WRDH should be saturated by the $W$-boson
predictions and then the competing a priori mixing contributions should be
suppressed necessarily by reducing the mixing angles. It is this situation that
will allow us to set a lower bound to the mass of mirror hadrons in the
manifest case.

Unfortunately, the mixing of ordinary and mirror hadrons cannot be derived
reliably starting at the quark level with a model, such as the one discussed by
Barr, Chang, and Senjanovi\'c in Ref.~\cite{barr}, due to our current inability
to perform QCD calculations in the non-perturbative regime of low energy
physics. One is therefore forced to introduce an ansatz to derive these a
priori mixing, as discussed in detail in Ref.~\cite{cinco}. Nevertheless, at
least for illustration purposes, mixing at the quark level can be shown to lead
to such an ansatz~\cite{rmf}.

We shall not repeat the details of the ansatz of Ref.~\cite{cinco}. Instead, we
shall discuss in more detail how the mixing angles between ordinary and mirror
hadrons appear in the physical hadrons (mass eigenstates) when the mass
matrices of hadrons are diagonalized. To be specific and to keep our analysis
simple, we shall discuss only the proton and $\Sigma^+$ system. This will be
sufficient for our purposes.

In this system the rotation matrix $R$ to get the physical ordinary $p_{\rm ph}$
and $\Sigma^+_{\rm ph}$ and mirror $\hat{p}_{\rm ph}$ and $\hat{\Sigma}^+_{\rm
ph}$ is a $4\times 4$ matrix. It can be split into the product of two $4\times
4$ matrices, $R=R^1R^0$. $R^0$ contains the large
Cabibbo-Kobayashi-Maskawa~\cite{ocho,nueve} angles and $R^1$ contains the
necessarily small mixing angles that will connect both the ordinary and mirror
hadrons. $R^0$ is block diagonal and it contains two non-zero $2\times2$
submatrices on its diagonal. The upper-left block actually contains only the
Cabibbo angle and operates to yield the flavor and parity eigenstates $p_{\rm
s}$ and $\Sigma^+_{\rm s}$. If no mirror hadrons exist at all this would be
reduced to two dimensions and the result would be

\begin{equation}
\left(
\begin{array}{cc}
{\bar p}_{\rm s} & {\bar \Sigma^+}_{\rm s}
\end{array}
\right)
\left(
\begin{array}{cc}
m^0_p & 0 \\
0 & m^0_{\Sigma^+}
\end{array}
\right)
\left(
\begin{array}{c}
p_{\rm s} \\
\Sigma^+_{\rm s}
\end{array}
\right)
\label{uno}
\end{equation}

The lower-right block operates analogously to yield the mirror flavor and parity
eigenstates $p_{\rm p}$ and $\Sigma^+_{\rm p}$ in a similar $2\times 2$ matrix.
So, in 4 dimensions the result of applying $R^0$ is

\begin{equation}
\left(
\begin{array}{cccc}
{\bar p}_{\rm s} & {\bar \Sigma^+}_{\rm s} &
{\bar p}_{\rm p} & {\bar \Sigma^+}_{\rm p}
\end{array}
\right)
\left(
\begin{array}{cccc}
m^0_p & 0 & 0 & 0 \\
0 & m^0_{\Sigma^+} & 0 & 0 \\
0 & 0 & {\hat m}^0_p & 0 \\
0 & 0 & 0 & {\hat m}^0_{\Sigma^+}
\end{array}
\right)
\left(
\begin{array}{c}
p_{\rm s} \\
\Sigma^+_{\rm s} \\
p_{\rm p} \\
\Sigma^+_{\rm p}
\end{array}
\right)
\label{dos}
\end{equation}

\noindent
The indices s and p stand for positive and negative parity, respectively, and
strong flavor is identified by the particle symbol. The case (\ref{dos})
contemplates ordinary and mirror matter still disconnected. When both worlds
are connected then the initial $4\times 4$ mass matrix contains two $2\times 2$
off-diagonal submatrices. In this case the action of $R^0$ yields

\begin{equation}
\left(
\begin{array}{cccc}
\bar{p}_{\rm s} & \bar{\Sigma}^+_{\rm s} &
{\bar p}_{\rm p} & \bar{\Sigma}^+_{\rm p}
\end{array}
\right)
\left(
\begin{array}{cccc}
m^0_p & 0 & \Delta_{11} & \Delta_{12} \\
0 & m^0_{\Sigma^+} & \Delta_{21} & \Delta_{22} \\
\Delta_{11} & \Delta_{21} & {\hat m}^0_p & 0 \\
\Delta_{12} & \Delta_{22} & 0 & {\hat m}^0_{\Sigma^+}
\end{array}
\right)
\left(
\begin{array}{c}
p_{\rm s} \\
\Sigma^+_{\rm s} \\
p_{\rm p} \\
\Sigma^+_{\rm p}
\end{array}
\right)
\label{tres}
\end{equation}

\noindent
instead of (\ref{dos}).
We are neglecting CP violation and therefore the $\Delta_{ij}$ can be taken as
real numbers. Notice that the zero entries in the two diagonal submatrices
still remain\footnote{This is a very important point. The CKM rotations cannot
lead to flavor and parity violation in strong and electromagnetic interactions.
So, for example, when these rotations are performed at the hadron level, as
initially proposed by Cabibbo, it is indispensable that the matrix containing
magnetic form factors remains diagonal along with the mass matrix. This is
possible because CKM rotations connect one kind of matter with itself only and
flavor and parity eigenstates can be defined after CKM rotations ($p_{\rm s}$,
etc.). This is not the case for $R^1$ and the reason is that it only connects
worlds of different kind.}. However, $R^0$ does affect the two off-diagonal
$2\times 2$ submatrices in the initial mass matrix. The effect of $R^0$ on them
is already incorporated in the $\Delta_{ij}$. The role of the rotation $R^1$ is
to finally diagonalize the full $4\times 4$ mass matrix of Eq.~(\ref{tres}) and
it is this final step that leads to the physical $p$ and $\Sigma^+$, which
contain flavor and parity mixing.

$R^1$ is in principle a complicated matrix with many angles.
However, one expects that the connection between the ordinary and the mirror
worlds be very small, because the mirror world would be far away (that is, with
very heavy masses).
Therefore, one must necessarily require the inequality

\begin{equation}
{\hat m}^0_p, {\hat m}^0_{\Sigma^+} \gg \Delta_{ij}, m^0_p, m^0_{\Sigma^+}
\label{cuatro}
\end{equation}

\noindent
This allows us then to keep $R^1$ to first order in the angles, namely,

\begin{equation}
(R^1)_{ij}\simeq\delta_{ij}+\epsilon_{ij}
\label{cinco}
\end{equation}

\noindent
where $\epsilon_{ij}=-\epsilon_{ji}$, $\delta_{ij}$ is the Kronecker delta, and
$i,j=1,\ldots,4$. There are only six relevant angles in $R^1$.

The action of $R^1$ upon the matrix $M$ in the sandwich of (\ref{tres}) leads
to a diagonal matrix $M_D$, i.\,e., $R^1M{R^1}^\dagger=M_D$, sandwiched between
the physical states and whose eigenvalues are the physical masses, namely,

\begin{equation}
\left(
\begin{array}{cccc}
\bar{p}_{\rm ph} & \bar{\Sigma}^+_{\rm ph} &
\bar{\hat{p}}_{\rm ph} & \bar{\hat{\Sigma}^+_{\rm ph}}
\end{array}
\right)
\left(
\begin{array}{cccc}
m_p & 0 & 0 & 0 \\
0 & m_{\Sigma^+} & 0 & 0 \\
0 & 0 & \hat{m}_p & 0 \\
0 & 0 & 0 & \hat{m}_{\Sigma^+}
\end{array}
\right)
\left(
\begin{array}{c}
p_{\rm ph} \\
\Sigma^+_{\rm ph} \\
\hat{p}_{\rm ph} \\
\hat{\Sigma}^+_{\rm ph}
\end{array}
\right)
\label{siete}
\end{equation}

\noindent
where

\begin{mathletters}
\label{ocho}
\begin{equation}
p_{\rm ph} =
p_{\rm s} + \epsilon_{12}\Sigma^+_{\rm s} + \epsilon_{13}p_{\rm p} +
\epsilon_{14}\Sigma^+_{\rm p}
\end{equation}
\begin{equation}
\Sigma^+_{\rm ph} =
\Sigma^+_{\rm s} - \epsilon_{12}p_{\rm s} + \epsilon_{23}p_{\rm p} +
\epsilon_{24}\Sigma^+_{\rm p}
\end{equation}
\end{mathletters}

\noindent
and analogous expressions for $\hat{p}_{\rm ph}$ and $\hat{\Sigma}^+_{\rm ph}$.
This diagonalization yields 16 equations.
Keeping the lowest relevant order in each equation, remembering that
${\hat m}^0_p$ and ${\hat m}^0_{\Sigma^+}$ are order zero and $\epsilon_{ij}$,
$\Delta_{ij}$, $m^0_p$ and $m^0_{\Sigma^+}$ are first order, one obtains
$m_p\simeq m^0_p$, $m_{\Sigma^+}\simeq m^0_{\Sigma^+}$,
${\hat m}_p\simeq {\hat m}^0_p$,
${\hat m}_{\Sigma^+}\simeq {\hat m}^0_{\Sigma^+}$, and

\begin{mathletters}
\label{ochon}
\begin{equation}
-m^0_p\epsilon_{12} + m^0_{\Sigma^+}\epsilon_{12} + \Delta_{21}\epsilon_{13} +
\Delta_{22}\epsilon_{14} + (\Delta_{11} + {\hat m}^0_p\epsilon_{13})\epsilon_{23} +
(\Delta_{12} + {\hat m}^0_{\Sigma^+}\epsilon_{14})\epsilon_{24} = 0
\label{ochona}
\end{equation}
\begin{equation}
-m^0_p\epsilon_{12} + m^0_{\Sigma^+}\epsilon_{12} + \Delta_{11}\epsilon_{23} +
\Delta_{12}\epsilon_{24} + (\Delta_{21} + {\hat m}^0_p\epsilon_{23})\epsilon_{13} +
(\Delta_{22} + {\hat m}^0_{\Sigma^+}\epsilon_{24})\epsilon_{14} = 0
\label{ochonb}
\end{equation}
\begin{equation}
\Delta_{11} + {\hat m}^0_p\epsilon_{13} = 0
\label{ochonc}
\end{equation}
\begin{equation}
\Delta_{12} + {\hat m}^0_{\Sigma^+}\epsilon_{14} = 0
\label{ochond}
\end{equation}
\begin{equation}
\Delta_{21} + {\hat m}^0_p\epsilon_{23} = 0
\label{ochone}
\end{equation}
\begin{equation}
\Delta_{22} + {\hat m}^0_{\Sigma^+}\epsilon_{24} = 0
\label{ochonf}
\end{equation}
\begin{equation}
-{\hat m}^0_p\epsilon_{34} + {\hat m}^0_{\Sigma^+}\epsilon_{34} = 0
\label{ochong}
\end{equation}
\end{mathletters}

\noindent
Eq.~(\ref{ochonb}) is just a rearrangement of Eq.~(\ref{ochona}), this
rearrangement will be useful for a later discussion. The remaining five
equations just repeat Eqs.~(\ref{ochonc}--\ref{ochong}). From all these
equations and still to lowest order, one obtains

\begin{mathletters}
\label{nueve}

\begin{equation}
|\epsilon_{12}| =
\left|
\frac{\Delta_{11}\epsilon_{23}+\Delta_{12}\epsilon_{24}}
{m_{\Sigma^+}-m_p}
\right|
\label{nuevea}
\end{equation}
 
\begin{equation}
|\epsilon_{12}| =
\left|
\frac{\Delta_{22}\epsilon_{14}+\Delta_{21}\epsilon_{13}}
{m_{\Sigma^+}-m_p}
\right|
\label{nueveb}
\end{equation}

\begin{equation}
|\epsilon_{13}| =
\left|
\frac{\Delta_{11}}{\hat{m}_p}
\right|
\label{nuevec}
\end{equation}

\begin{equation}
|\epsilon_{14}| =
\left|
\frac{\Delta_{12}}{\hat{m}_{\Sigma^+}}
\right|
\label{nueved}
\end{equation}

\begin{equation}
|\epsilon_{23}| =
\left|
\frac{\Delta_{21}}{\hat{m}_p}
\right|
\label{nuevee}
\end{equation}

\begin{equation}
|\epsilon_{24}| =
\left|
\frac{\Delta_{22}}{\hat{m}_{\Sigma^+}}
\right|
\label{nuevef}
\end{equation}
\end{mathletters}

\noindent
and $\epsilon_{34}\simeq 0$.
The ordering of these equations corresponds to the ordering of
Eqs.~(\ref{ochon}).

We shall need absolute values only.
The angles $\epsilon_{12}$, $\epsilon_{14}$, and $\epsilon_{23}$ may give
observable effects in NLDH and WRDH.
Their values were obtained in Refs.~\cite{cinco} and
\cite{seis}, assuming that the mixings of Eqs.~(\ref{ocho}) give contributions
that saturate the corresponding NLDH and WRDH available data.
These are the most favorable conditions to observe the mixing with mirror
hadrons.
The contributions of the $W$-boson were assumed not to be enhanced, i.\,e., its
$\Delta I=1/2$ contributions were assumed to be at the same level of its
$\Delta I=3/2$ contributions and, accordingly, were neglected.

The magnitudes of the angles obtained are

\begin{equation}
|\epsilon_{12}|=(4.9\pm 2.0)\times 10^{-6}
\label{doce}
\end{equation}

\begin{equation}
|\epsilon_{14}|=(0.22\pm 0.09)\times 10^{-6}
\label{trece}
\end{equation}

\begin{equation}
|\epsilon_{23}|=(0.26\pm 0.09)\times 10^{-6}
\label{catorce}
\end{equation}

\noindent
In these last two references these angles were identified as $\sigma$, $\delta$,
and $\delta'$, respectively.

Notice that $|\epsilon_{12}|$ is an order of magnitude larger than
$|\epsilon_{14}|$ and $|\epsilon_{23}|$.
The angles $|\epsilon_{13}|$ and $|\epsilon_{24}|$ have not been determined.
However, we expect them to be at most of the same order of magnitude as the
first three.
To obtain a lower bound on mirror baryon masses we must assume that
$|\epsilon_{13}|$ and/or $|\epsilon_{24}|$ are of the order of magnitude of
$|\epsilon_{12}|$ namely, $|\epsilon_{13}|\simeq|\epsilon_{12}|$ and/or
$|\epsilon_{24}|\simeq|\epsilon_{12}|$.
Looking back at Eqs.~(\ref{nuevea}) and (\ref{nueveb}) we may conclude then
that

\begin{equation}
\frac{|\Delta_{ab}|}{m_{\Sigma^+}-m_p} \simeq 1
\label{quince}
\end{equation}

\noindent
where the pair of indices ${ab}$, take the values 12 or 21.
To get Eq.~(\ref{quince}) notice that the first summands in
Eqs.~(\ref{nuevea}) and (\ref{nueveb})
involve the smaller angles $|\epsilon_{14}|$ and $|\epsilon_{23}|$
of Eqs.~(\ref{trece}) and (\ref{catorce}). If these summands were to dominate
the numerators of Eqs.~(\ref{nuevea}) and (\ref{nueveb}) then the indices
${ab}$ would take the values 11 or 22 and the right hand side of (\ref{quince})
would become 10. This option leads to a higher lower bound and we discard it,
accordingly.

Using the value of the mass difference $m_{\Sigma^+}-m_p\simeq 0.25{\rm GeV}$,
Eq.~(\ref{quince}) gives $|\Delta_{12}|\simeq 0.25{\rm GeV}$.
Substituting this into Eq.~(\ref{nueved}) and using the central value of
Eq.~(\ref{catorce}) we obtain an order of magnitude lower bound for $\hat{m}_p$,

\begin{equation}
\hat{m}_p \geq {\cal O} (10^6 {\rm GeV}).
\label{dieciocho}
\end{equation}

\noindent
From Eq.~(\ref{nuevee}) a similar bound is obtained.

At this point it is important to emphasize in what sense Eq.~(\ref{dieciocho})
is to be understood as a lower bound.
So far, it is possible to assume that the $\Delta_{ij}$ of Eq.~(\ref{quince})
may become smaller, to the extent $\Delta_{ij}\to 0$.
Then the mirror and the ordinary baryons become decoupled.
Of course, once decoupled one cannot set any sort of lower bound on mirror
baryon masses using information of the ordinary baryon world.
This situation corresponds to requiring that mirror matter does not give
observable effects in our ordinary matter world.
Eq.~(\ref{dieciocho}) is to be understood in the opposite sense.
That is, what values of the masses of mirror baryons would lead to observable
effects in our world?.
This clearly requires $\Delta_{ij}\neq 0$ as used above and then
Eq.~(\ref{dieciocho}) tells us that a value of below the bound of
Eq.~(\ref{dieciocho}) may lead to effects in NLDH and WRDH which exceed the
level experimentally observed for these decays, even the equality sign exceeds
the level predicted by the $W$ boson when the observed enhancement of its
$\Delta I = 1/2$ contributions is assumed to arise within the minimal standard
model.
It is in this sense that Eq.~(\ref{dieciocho}) represents a lower bound.

To our knowledge this is the only available lower bound on the mass of mirror
hadrons.
Other bounds on mirror matter refer to the values of their mixing angle with
ordinary matter.
Such bounds, from precision tests of the standard model, were thoroughly discussed
by Langacker, Luo, and Mann in Ref.~\cite{once}, but no attempt was made there
to get bounds on masses of mirror fermions.
The bounds for the mixing angles obtained there are around $3\times 10^{-2}$, which
are much too high compared with Eqs.~(\ref{doce})--(\ref{catorce}).

Besides the (rather trivial) way to avoid the lower bound of Eq.~(\ref{dieciocho})
that we just discussed, one other way to avoid it is by relaxing the manifest
mirror symmetry assumption that we made.
Our bound depends crucially on the assumption that mirror matter shares the same
strong and electromagnetic (e.m.) interactions with ordinary matter.
The question arises then if it is possible to keep such manifest mirror symmetry
while making mirror matter much heavier than ordinary matter.
We cannot give a rigorous answer to this question.
However, symmetry breaking scenarios are conceivable that allow breaking mirror
symmetry without affecting the strong and e.m.\ effective couplings in the
mirror sector.
The model of Ref.~\cite{barr} provides useful guidance in this respect.
This model is conceived at the quark level.
But, as we discussed in Ref.~\cite{rmf} our phenomenological group-theoretic
ansatz of Ref.~\cite{cinco} can be qualitatively derived from a quark level
approach.
Using the model of Ref.~\cite{barr} one can show that after breaking mirror
(in this case left-right) symmetry one can reconstruct the e.m.\ current
operator of physical quarks as a proper flavor-conserving four-vector which
couples to ordinary and mirror physical matter with common charges.
Also, since the QCD interactions are described by the same $\rm{SU}(3)_{\rm{C}}$
in direct product with the electroweak sector, the strong interactions of
mirror physical hadrons remain the same ones of ordinary hadrons, after mirror
matter was made much heavier. Therefore, the answer to the previous question is
qualitatively in the affirmative.

The lower bound of Eq.~(\ref{dieciocho}) is impressively high.
It means that producing mirror matter on earth is way far into the future.
Even if one is willing to abandon the manifest mirror symmetry assumption (and
allowing mirror masses to become smaller) would not help, because the coupling
through strong and e.m. interactions to ordinary matter would be greatly
reduced and, since machines on earth would be made out of ordinary matter, it
would still be very difficult to produce mirror matter with them. Also, the
bound of Eq.~(\ref{dieciocho}) means that mirror matter may not be a very good
candidate for dark matter in the universe~\cite{zurab}, although there always
exists the possibility of detecting it in cosmic rays. Whatever the real
situation may be, the possibility exists that mirror matter may give unwanted
effects in low energy physics if it is too light, and this is what makes
(\ref{dieciocho}) a lower bound.

We would like to thank CONACyT (Mexico) for
partial support.

\end{document}